\documentclass[letterpaper, 10 pt, conference]{ieeeconf}  

\IEEEoverridecommandlockouts                              
\usepackage{amsmath}
\usepackage{graphics}
\usepackage{epsfig} 
\usepackage{mathptmx} 
\usepackage{times} 
\usepackage{amsmath} 
\usepackage{amssymb}  

\title{\LARGE \bf 
RNN Decoding of Linear Block Codes}

\author{Eliya~Nachmani$^{1}$, Elad~Marciano$^{2}$, David~Burshtein$^{3}$ and Yair~Be'ery$^{4}$ %
\thanks{$^{1}$Eliya Nachmani is with the School of Electrical Engineering, Tel-Aviv University,
{\tt\small enk100@gmail.com}}%
\thanks{$^{2}$Elad Marciano is with the School of Electrical Engineering, Tel-Aviv University,
{\tt\small eladmarc@gmail.com}}%
\thanks{$^{3}$David~Burshtein is with the School of Electrical Engineering, Tel-Aviv University,
{\tt\small burstyn@eng.tau.ac.il}}%
\thanks{$^{4}$Yair~Be'ery is with the School of Electrical Engineering, Tel-Aviv University,
{\tt\small ybeery@eng.tau.ac.il}}%
}

\begin{document}



\maketitle
\thispagestyle{empty}
\pagestyle{empty}

\begin{abstract}
Designing a practical, low complexity, close to optimal, channel decoder for powerful algebraic codes with short to moderate block length is an open research problem. Recently it has been shown that a feed-forward neural network architecture can improve on standard belief propagation decoding, despite the large example space. In this paper we introduce a recurrent neural network architecture for decoding linear block codes. Our method shows comparable bit error rate results compared to the feed-forward neural network with significantly less parameters. We also demonstrate improved performance over belief propagation on sparser Tanner graph representations of the codes. Furthermore, we demonstrate that the RNN decoder can be used to
improve the performance or alternatively reduce the computational complexity of the mRRD algorithm for low complexity, close to optimal, decoding of short BCH codes.
\end{abstract}

\section{INTRODUCTION}

In recent years deep learning methods have demonstrated significant improvements in various tasks. These methods outperform human-level object detection in some tasks \cite{resnet}, and achieve state-of-the-art results in machine translation \cite{nmt} and speech processing \cite{graves2013speech}. Additionally, deep learning combined with reinforcement learning techniques was able to beat human champions in challenging games such as Go \cite{d_silver}. 

Error correcting codes for channel coding are used in order to enable reliable communications at rates close to the Shannon capacity. A well-known family of linear error correcting codes are the linear low-density parity-check (LDPC) codes \cite{galmono}. LDPC codes achieve near Shannon channel capacity with the belief propagation (BP) decoding algorithm, but can typically do so for relatively large block lengths.
For short to moderate high density parity check (HDPC) codes \cite{jiang2006iterative,dimnik2009improved,yufit2011efficient,zhang2012adaptive,helmling2014efficient}, such as common powerful linear algebraic codes, the regular BP algorithm obtains poor results compared to the optimal maximum likelihood decoder. On the other hand, the importance of close to optimal low complexity, low latency and low power decoders of short to moderate codes has grown with the emergence of applications driven by the Internet of Things.

Recently in \cite{nachmani} it has been shown that deep learning methods can improve the BP decoding of HDPC codes using a neural network. They formalized the belief propagation algorithm as neural network and showed that it can improve the decoding by $0.9{\rm dB}$ in the high SNR regime. A key property of the method is that it is sufficient to train the neural network decoder using a single codeword (e.g., the all-zero codeword), since the architecture guarantees the same error rate for any chosen transmitted codeword.

Later, Lugosch \& Gross \cite{lugosch} proposed an improved neural network architecture that achieves similar results to \cite{nachmani} with less parameters and reduced complexity. The main difference was that they use the min-sum algorithm instead of the sum-product algorithm. Gruber et al. \cite{tenbrink} proposed a neural net decoder with an unconstrained graph (i.e., fully connected network) and show that the network gets close to maximum likelihood results for very small block codes, $N=16$. Also, O’Shea \& Hoydis \cite{AutoencoderComm} proposed to use an autoencoder as a communication system for small block code with $N=7$.

In this work we modify the architecture of \cite{nachmani} to a recurrent neural network (RNN) and show that it can achieve up to $1.5{\rm dB}$ improvement over the belief propagation algorithm in the high SNR regime. The advantage over the feed-forward architecture of \cite{nachmani} is that it reduces the number of parameters.
We also investigate the performance of the RNN decoder on parity check matrices with lower densities and fewer short cycles and show that despite the fact that we start with reduced cycle matrix, the network can improve the performance up to $1.0{\rm dB}$. The output of the training algorithm can be interpreted as a soft Tanner graph that replaces the original one. State of the art decoding algorithms of short to moderate algebraic codes, such as \cite{fossOSD1,dimnik2009improved,helmling2014efficient}, utilize the BP algorithm as a component in their solution. Thus, it is natural to replace the standard BP decoder with our trained RNN decoder, in an attempt to improve either the decoding performance or its complexity. In this work we demonstrate, for a BCH(63,36) code, that such improvements can be realized by using the RNN decoder in the mRRD algorithm.

\section{Related Work} 
\subsection{Belief Propagation Algorithm}
The BP decoder \cite{galmono,ru_book} is a messages passing algorithm. The algorithm is constructed from the Tanner graph which is a graphical representation of the parity check matrix. The graphical representation consists of edges and nodes. There are two type of nodes:
\begin{itemize}
\item Check nodes - corresponding to rows in the parity check matrix.
\item Variable nodes - corresponding to columns in the parity check matrix.
\end{itemize}

The edges correspond to ones in the parity check matrix. 
The messages are transmitted over edges. Consider a code with block length $N$. The input to the algorithm is a vector of size $N$, that consists of the log-likelihood ratios (LLRs) of the channel outputs. We consider an algorithm with $L$ decoding iterations. The LLR values, $v=1,2,\ldots,N$, are given by
$$
l_v = \log\frac{\Pr\left(C_v=1 | y_v\right)}{\Pr\left(C_v=0 | y_v\right)}
$$
where $y_v$ is the channel output corresponding to the $v$th codebit, $C_v$. 
The $L$ iterations of the BP decoder are represented in \cite{nachmani} using the following trellis graph. The input layer consists of $l_v, v=1,...,N$ nodes. The following $2L$ layers in the graph have size $E$ each, where $E$ is the number of edges in the Tanner graph (number of ones in the parity check matrix). The last layer has size $N$, which is the length of the code.

The messages transmitted over the trellis graph are the following. Consider hidden layer $i$, $i=1,2,\ldots,2L$, and let $e=(v,c)$ be the index of some processing element in that layer. We denote by $x_{i,e}$, the output message of this processing element. For odd (even, respectively), $i$, this is the message produced by the BP algorithm after $\lfloor (i-1)/2 \rfloor$ iterations, from variable to check (check to variable) node.

For odd $i$ and $e=(v,c)$ we have (recall that the self LLR message of $v$ is $l_v$),
\begin{equation}
x_{i,e=(v,c)} = l_v + \sum_{e'=(c',v),\: c'\ne c} x_{i-1,e'}
\label{eq:x_ie_RB}
\end{equation}
under the initialization, $x_{0,e'}=0$ for all edges $e'$ (in the beginning there is no information at the parity check nodes). The summation in~\eqref{eq:x_ie_RB} is over all edges $e'=(c',v)$ with variable node $v$ except for the target edge $e=(c,v)$. Recall that this is a fundamental property of message passing algorithms ~\cite{ru_book}.

Similarly, for even $i$ and $e=(c,v)$ we have,
\begin{equation}
x_{i,e=(c,v)} = 2\tanh^{-1} \left( \prod_{e'=(v',c),\: v'\ne v} \tanh \left( \frac{x_{i-1,e'}}{2} \right) \right)
\label{eq:x_ie_LB}
\end{equation}

The final $v$th output of the network is given by
\begin{equation}
o_v = l_v + \sum_{e'=(c',v)} x_{2L,e'}
\label{eq:ov}
\end{equation}
which is the final marginalization of the BP algorithm.

\subsection{Neural Sum Product Algorithm}
Nachmani et al. \cite{nachmani} have suggested a parameterized deep neural network decoder as a generalization of the BP algorithm. They use the trellis representation of the BP algorithm with weights associated with each edge of the Tanner graph. These weights are trained with stochastic gradient descent. More precisely, the equations to the neural sum product algorithm are -
\begin{align}
&x_{i,e=(v,c)} =\nonumber \\ 
&=\tanh \Biggl( \frac{1}{2} \Biggr.\Biggl(w_{i,v} l_v + \sum_{e'=(c',v),\: c'\ne c} w_{i,e,e'} x_{i-1,e'}\Biggr) \Biggl. \Biggr ) 
\label{eq:x_ie_RB_NN}
\end{align}

for odd $i$,
\begin{equation}
x_{i,e=(c,v)} = 2\tanh^{-1} \left( \prod_{e'=(v',c),\: v'\ne v}{x_{i-1,e'}}\right)
\label{eq:x_ie_LB_NN}
\end{equation}
for even $i$, and
\begin{equation}
o_v = \sigma \left( w_{2L,v} l_v + \sum_{e'=(c',v)} w_{2L,v,e'} x_{2L,e'} \right)
\label{eq:ov_NN}
\end{equation}
where $\sigma(x) \equiv \left( 1+e^{-x} \right)^{-1}$ is a sigmoid function. This algorithm coincides with the BP algorithm if all the weights are set to one (except for the sigmoid function at the output). Therefore the neural sum product algorithm cannot be inferior to the plain BP algorithm.

\begin{figure}[thpb]
	\centering
    \includegraphics[width=0.983101925\linewidth]{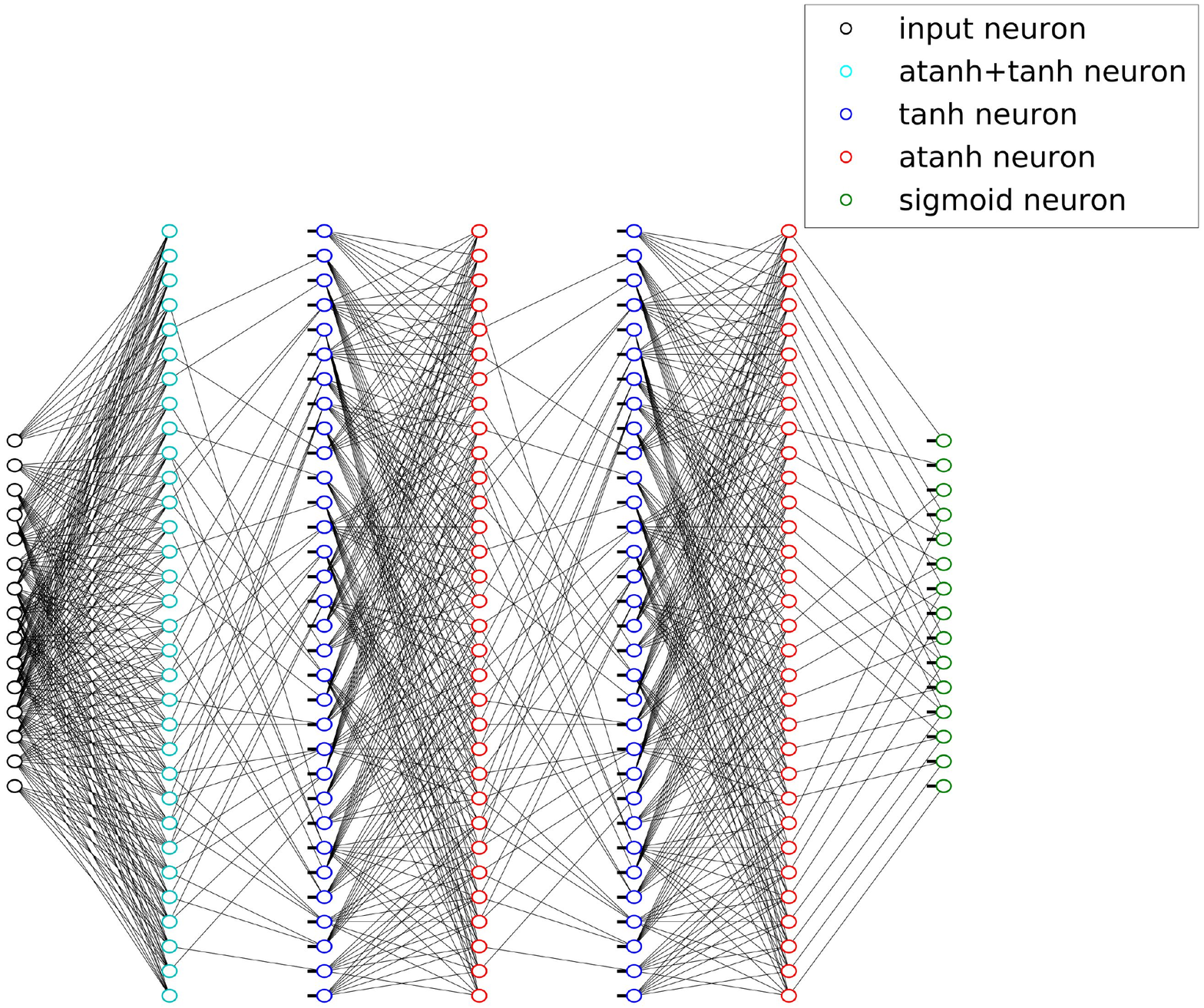}
	\caption{Deep Neural Network Architecture For BCH(15,11) with 5 hidden layers which correspond to 3 full BP iterations.}
	\label{fig:BCH_15_11_arch}
\end{figure}

The neural sum product algorithm satisfies the message passing symmetry conditions \cite{ru_book}[Definition 4.81]. Therefore the error rate is independent of the transmitted codeword. As a result the network can be trained by using noisy versions of a single codeword.
The time complexity of the neural sum product algorithm is similar to plain BP algorithm. However, the neural sum product algorithm requires more multiplications and parameters then the plain BP algorithm. The neural network architecture is illustrated in Figure~\ref{fig:BCH_15_11_arch} for a BCH(15,11) code.

\subsection{Modified Random Redundant Iterative (mRRD) Algorithm}
\begin{figure}[thpb]
	\centering  \includegraphics[width=0.85\linewidth]{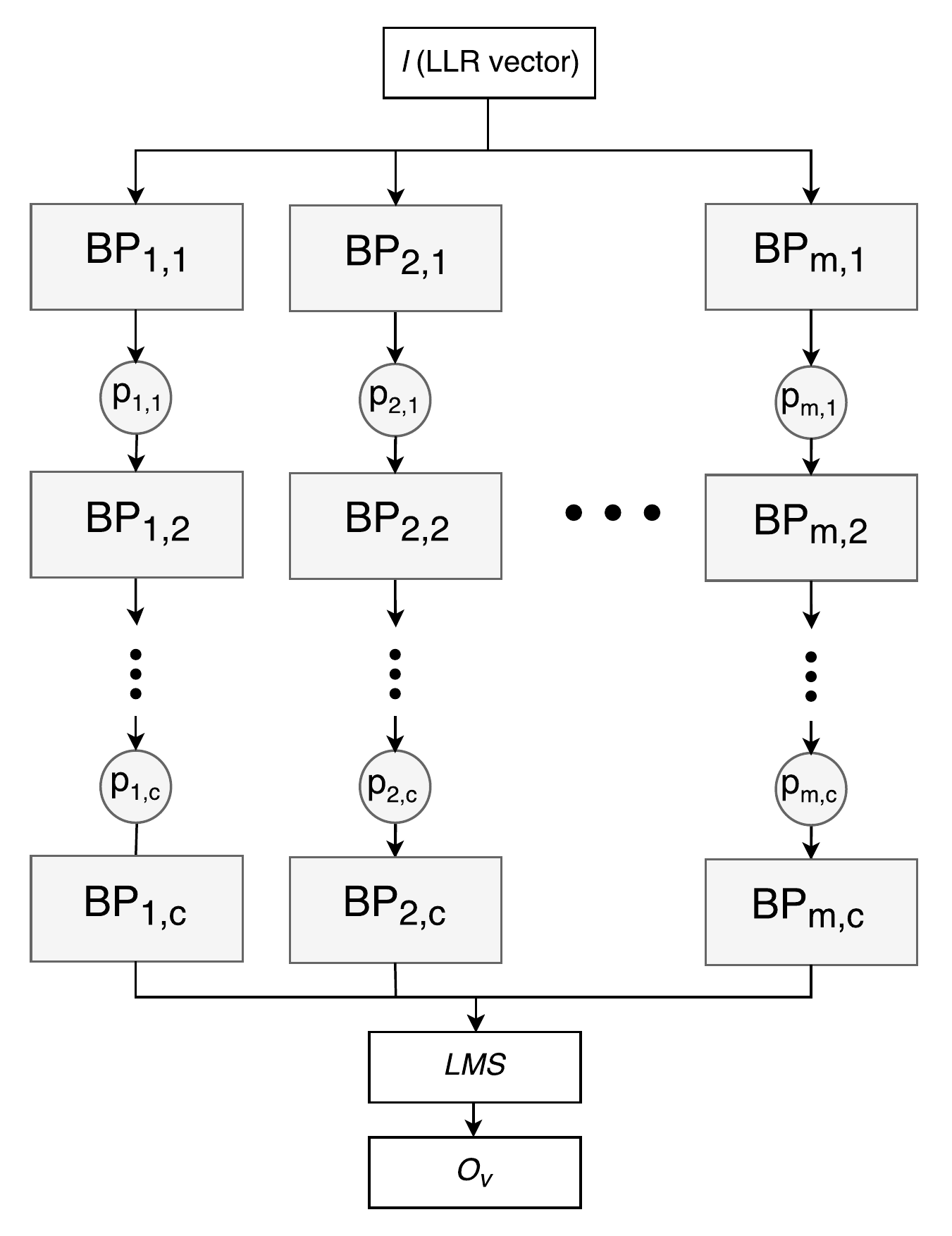}
	\caption{mRRD decoder with $m$ parallel iterative decoders, and $c$ BP blocks in each iterative decoder. The circles represent random permutations from the Automorphism Group of the code.}
	\label{fig:mrrd_diag}
\end{figure}

Dimnik and Be'ery \cite{dimnik2009improved} proposed an iterative algorithm for decoding HDPC codes based on the RRD \cite{cycle_reduce} and the MBBP \cite{mbbp} algorithms. The mRRD algorithm is a close to optimal low complexity decoder for short length ($N<100$) algebraic codes such as BCH codes. This algorithm uses $m$ parallel decoder branches, each comprising of $c$ applications of several (e.g. 2) BP decoding iterations, followed by a random permutation from the Automorphism Group of the code, as shown in Figure~\ref{fig:mrrd_diag}. The decoding process in each branch stops if the decoded word is a valid codeword. The final decoded word is selected with a least metric selector (LMS) as the one for which the channel output has the highest likelihood. More details can be found in \cite{dimnik2009improved}.

\section{Methods} 
\subsection{BP-RNN Decoding}
\begin{figure}[thpb]
	\centering
    \includegraphics[width=0.953101925\linewidth]{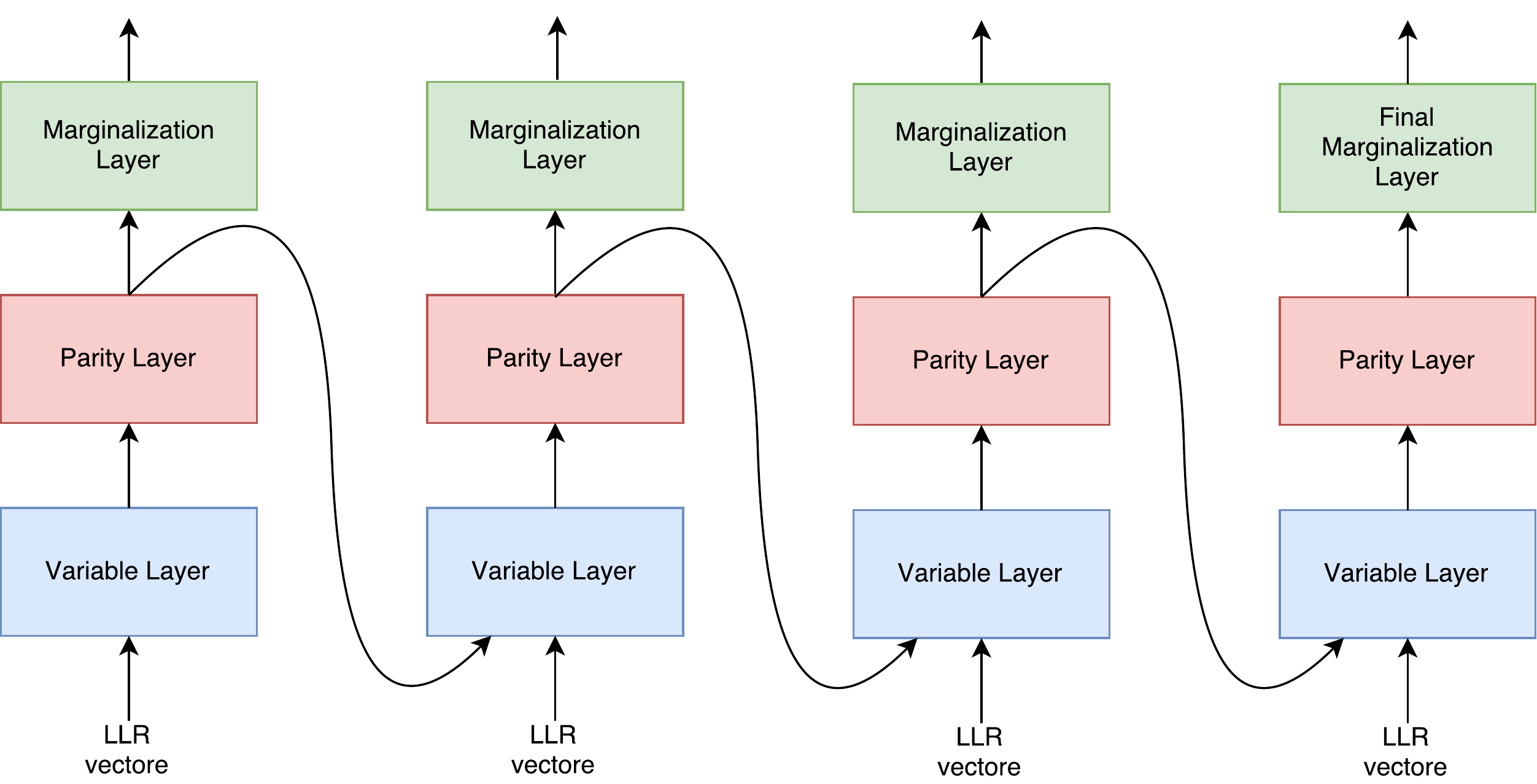}
	\caption{Recurrent Neural Network Architecture with unfold 4 which correspond to 4 full BP iterations.}
	\label{fig:rnn_unfold}
\end{figure}

We suggest the following parameterized deep neural network decoder which is a constrained version of the BP decoder of the previous section. We use the same trellis representation as in \cite{nachmani} for the decoder. The difference is that now the weights of the edges in the Tanner graph are tied, i.e. they are set to be equal in each iteration. This tying transfers the feed-forward architecture of \cite{nachmani} into a recurrent neural network architecture. More precisely, the equations of the proposed architecture for time step $t$ are
\begin{align}
&x_{i,e=(v,c)} = \nonumber \\ 
&= \tanh \Biggl( \frac{1}{2}  \Biggr. \Biggl(w_{v} l_v + \sum_{e'=(c',v),\: c'\ne c} w_{e,e'} x_{t-1,e'}\Biggr) \Biggl. \Biggr) 
\label{eq:x_ie_RB_NN_rnn}
\end{align}

\begin{equation}
x_{t,e=(c,v)} = 2\tanh^{-1} \left( \prod_{e'=(v',c),\: v'\ne v}{x_{t,e'}}\right)
\label{eq:x_ie_LB_NN_rnn}
\end{equation}
For time step, $t$, we have
\begin{equation}
o_v = \sigma \left( w_{t,v} l_v + \sum_{e'=(c',v)} w_{t,v,e'} x_{t,e'} \right)
\label{eq:ov_NN}
\end{equation}
where $\sigma(x) \equiv \left( 1+e^{-x} \right)^{-1}$ is a sigmoid function. We initialize the algorithm by setting $x_{0,e}=0$ for all $e=(c,v)$. The proposed architecture also preserves the symmetry conditions. As a result the network can be trained by using noisy versions of a single codeword. The training is done as before with a cross
entropy loss function at the last time step - 
\begin{equation}
L{(o,y)}=-\frac{1}{N}\sum_{v=1}^{N}y_{v}\log(o_{v})+(1-y_{v})\log(1-o_{v})
\label{eq:cross_entropy}
\end{equation} 
where $o_{v}$, $y_{v}$ are the final deep recurrent neural network output and the actual $v$th component of the transmitted codeword.
The proposed recurrent neural network architecture has the property that after every time step we can add final marginalization and compute the loss of these terms using ~\eqref{eq:cross_entropy}. Using multiloss terms can increase the gradient update at the backpropagation through time algorithm and allow learning the earliest layers. At each time step we add the final marginalization to loss:
\begin{equation}
L{(o,y)}=-\frac{1}{N}\sum_{t=1}^{T}\sum_{v=1}^{N}y_{v}\log(o_{v,t})+(1-y_{v})\log(1-o_{v,t})
\label{eq:multiloss_cross_entropy}
\end{equation}
where $o_{v,t}$, $y_{v}$ are the deep neural network output at the time step $t$ and the actual $v$th component of the transmitted codeword. This network architecture is illustrated in Figure~\ref{fig:rnn_unfold}. Nodes in the variable layer implement~\eqref{eq:x_ie_RB_NN_rnn}, while nodes in the parity layer implement~\eqref{eq:x_ie_LB_NN_rnn}. Nodes in the marginalization layer implement~\eqref{eq:ov_NN}. The training goal is to minimize~\eqref{eq:multiloss_cross_entropy}.

\subsection{mRRD-RNN Decoding}
We propose to combine the BP-RNN decoding algorithm with the mRRD algorithm. We can replace the BP blocks in the mRRD algorithm with our BP-RNN decoding scheme. The proposed mRRD-RNN decoder algorithm should achieve near maximum likelihood performance with less computational complexity.

\section{Experiments And Results} 
\subsection{BP-RNN}
We apply our method to different linear codes, BCH(63,45), BCH(63,36), BCH(127,64) and BCH(127,99). In all experiments the results of training, validation and test sets are identical, we did not observe overfitting in our experiments. Details about our experiments
and results are as follows. It should be noted that we have not trained the parameters $w_v$ in ~\eqref{eq:x_ie_RB_NN_rnn}, i.e. we set $w_v =1$

Training was conducted using stochastic gradient descent with mini-batches. The training data is created by transmitting the zero codeword through an AWGN channel with varying SNRs ranging from $1{\rm dB}$ to $8{\rm dB}$. The mini-batch size was $120$, $80$ and $40$ examples to BCH codes with $N=63$, BCH(127,99) and BCH(127,64) respectively. We applied the RMSPROP~\cite{rmsprop} rule with a learning rate equal to $0.001$, $0.0003$ and $0.003$ to BCH codes with $N=63$, BCH(127,99) and BCH(127,64) respectively. The neural network has $2$ hidden layers at each time step, and unfold equal to $5$ which corresponds to $5$ full iterations of the BP algorithm. 
At test time, we inject noisy codewords after transmitting through an AWGN channel and measure the bit error rate (BER) in the decoded codeword at the network output.
The input $x_{t-1,e’}$ to ~\eqref{eq:x_ie_RB_NN_rnn} is clipped such that the absolute value of the input is always smaller than some positive constant $A < 10$. This is also required for a practical implementation of the BP algorithm. \\

\subsubsection{BER For BCH With $N=63$} 

\begin{figure}[thpb]
	\centering	\includegraphics[width=1.1\linewidth]{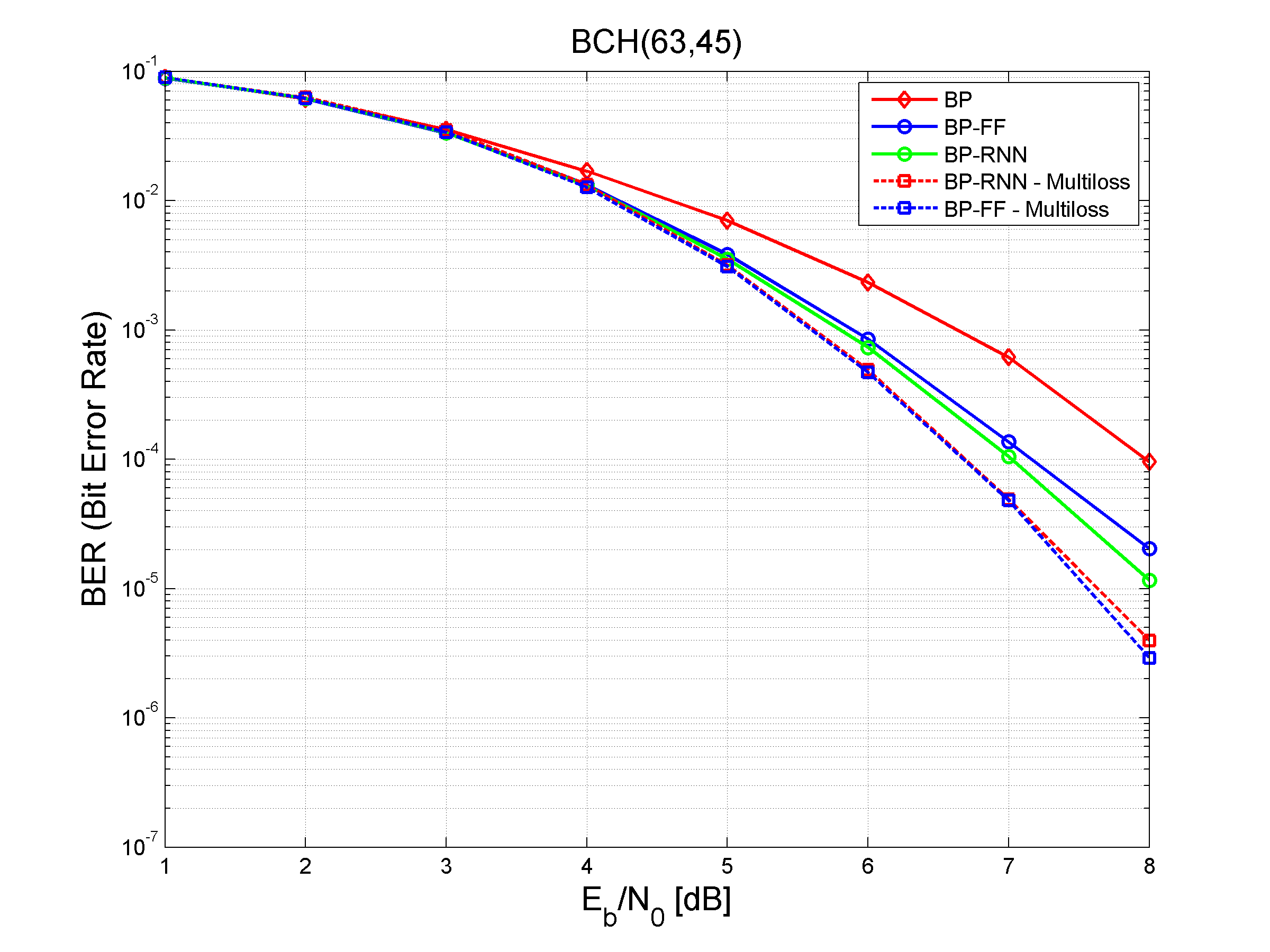}
	\caption{BER results for BCH(63,45) code trained with regular parity check matrix}
	\label{fig:bch_63_45_ber_regular}
\end{figure}

\begin{figure}[thpb]
	\centering	\includegraphics[width=1.1\linewidth]{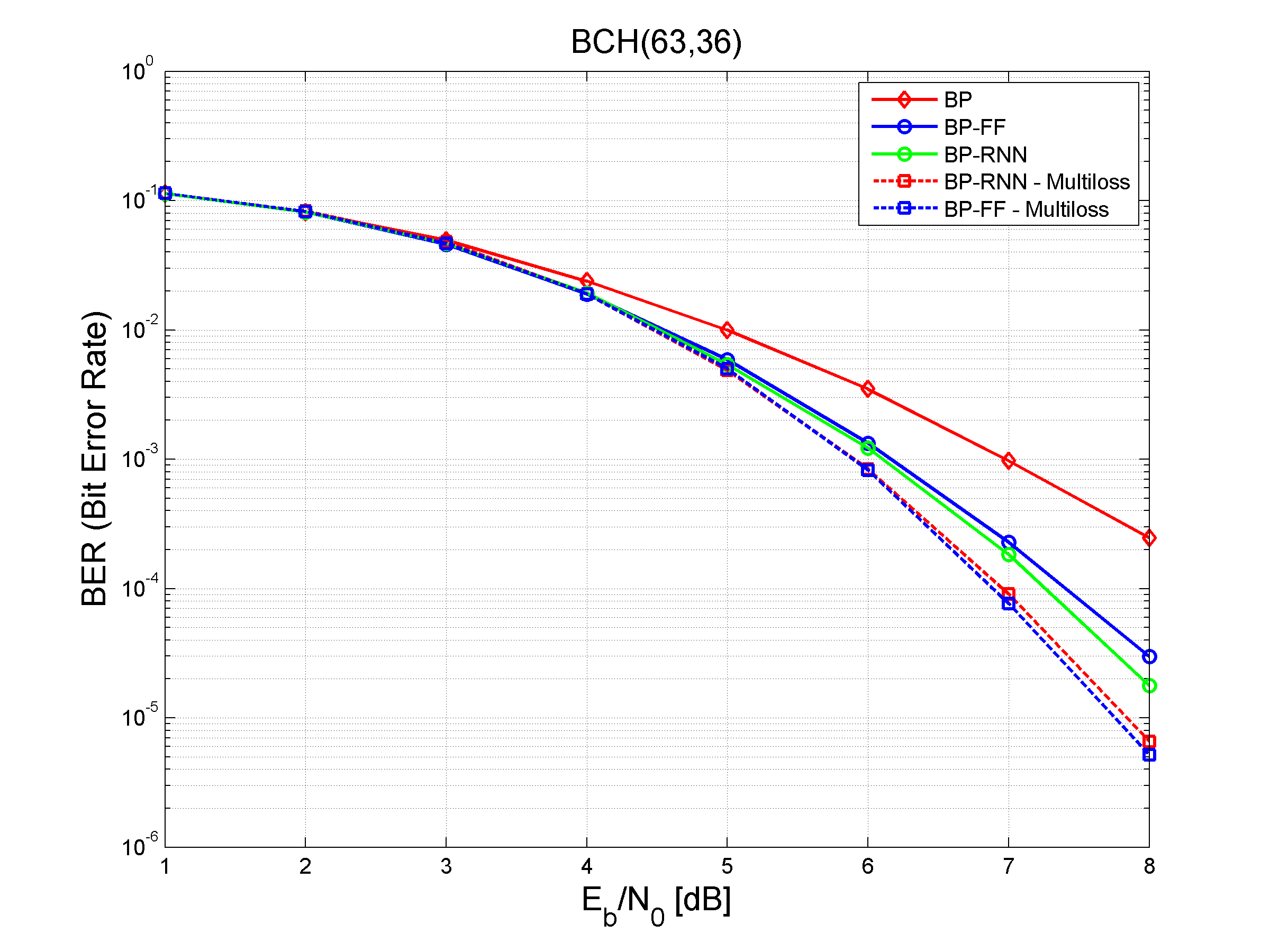}
	\caption{BER results for BCH(63,36) code trained with regular parity check matrix}
	\label{fig:bch_63_36_ber_regular}
\end{figure}

\hfill \break \newline In Figures~\ref{fig:bch_63_45_ber_regular},~\ref{fig:bch_63_36_ber_regular}, we provide the bit-error-rate for BCH code with $N=63$ for regular parity check matrix based on \cite{parity_g}. As can be seen from the figures, the BP-RNN decoder outperforms the BP feed-forward (BP-FF) decoder by $0.2{\rm dB}$. Not only that we improve the BER the network has less parameters. Moreover, we can see that the BP-RNN decoder obtains comparable results to the BP-FF decoder when training with multiloss. Furthermore, for the BCH(63,45) and BCH(63,36) there is an improvement up to $1.3{\rm dB}$ and $1.5{\rm dB}$, respectively, over the plain BP algorithm.

\begin{figure}[thpb]
	\centering	\includegraphics[width=1.1\linewidth]{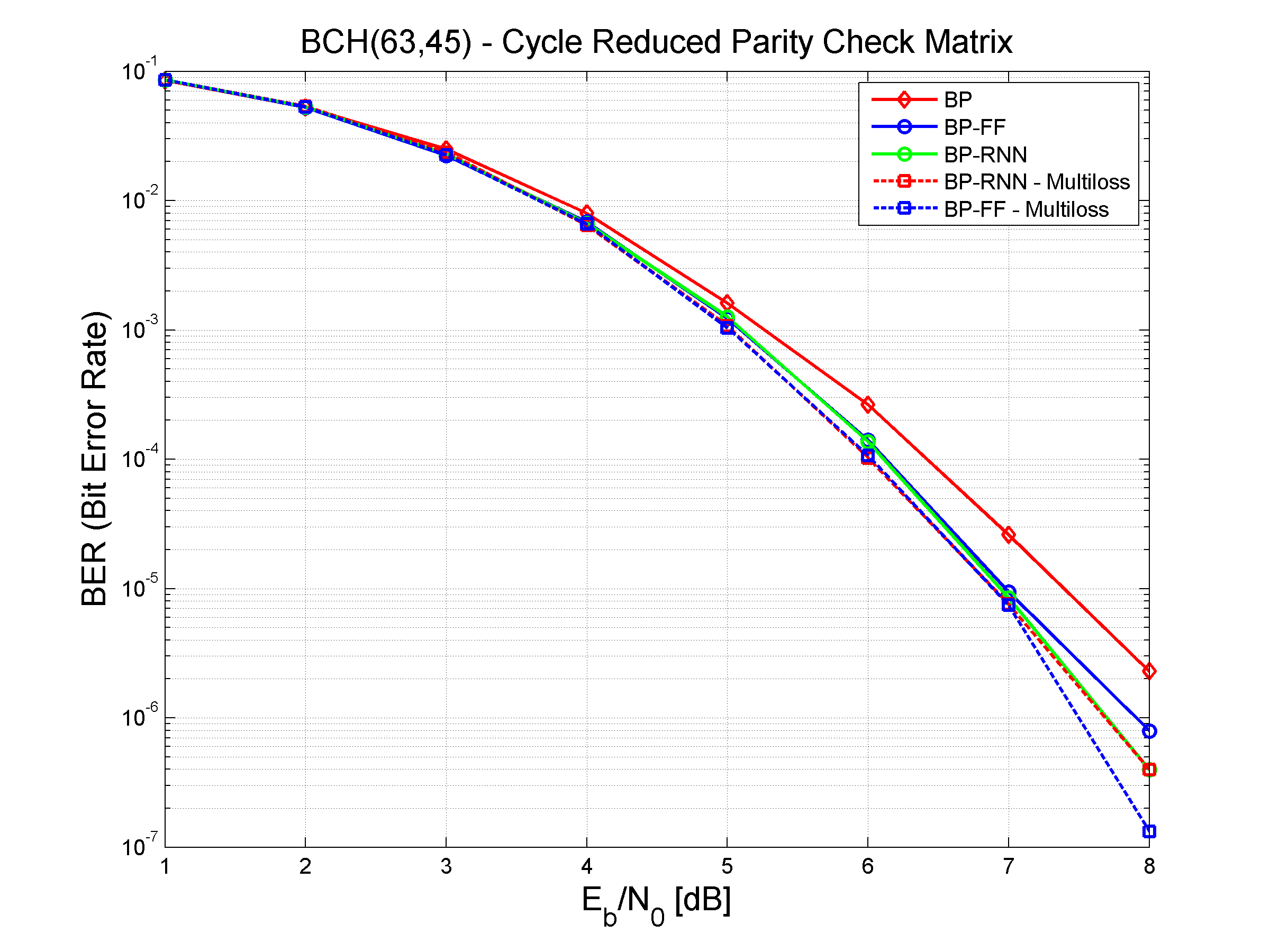}
	\caption{BER results for BCH(63,45) code trained with cycle reduced parity check matrix}
	\label{fig:bch_63_45_ber_iregular}
\end{figure}

\begin{figure}[thpb]
	\centering	\includegraphics[width=1.1\linewidth]{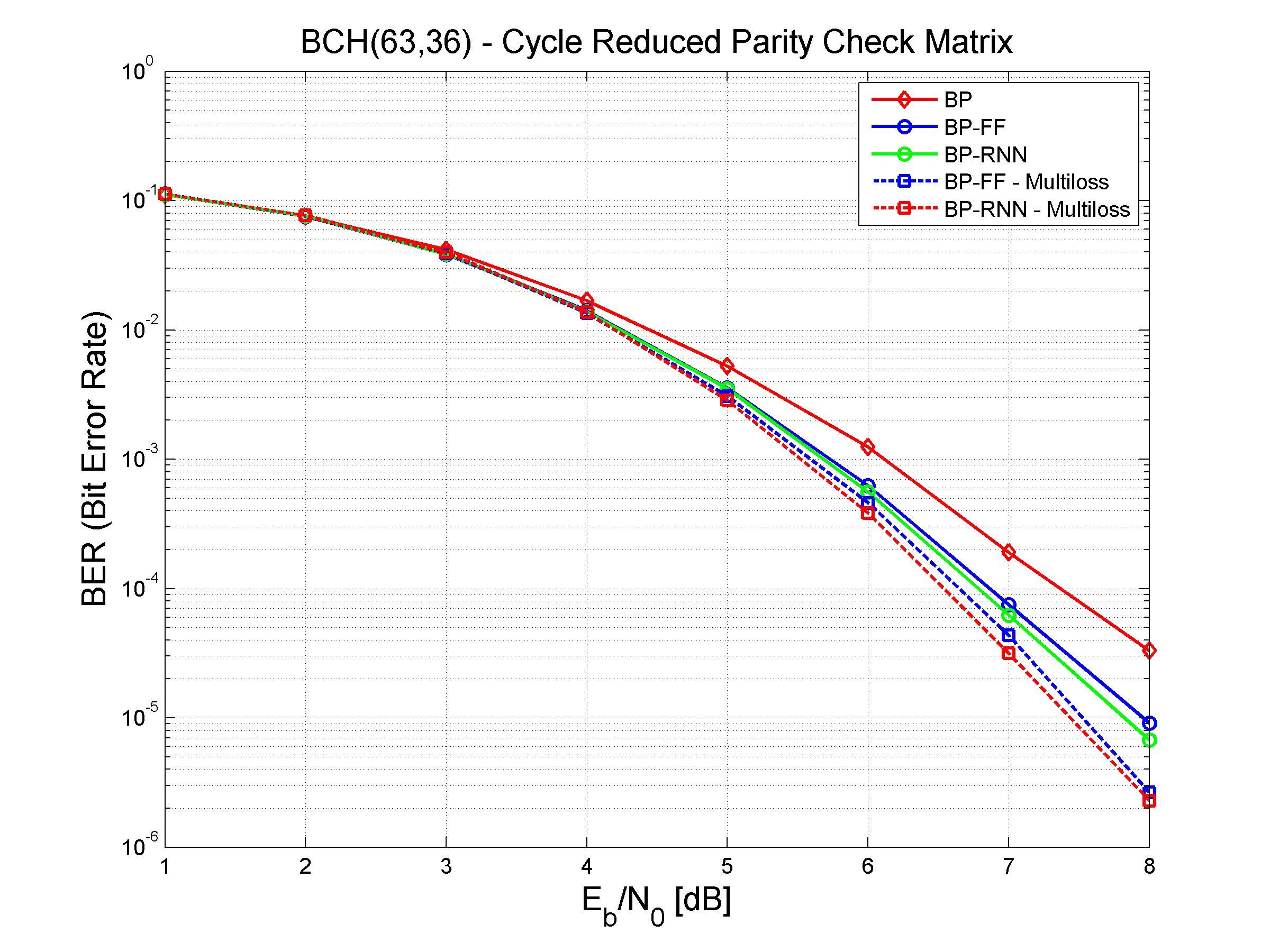}
	\caption{BER results for BCH(63,36) code trained with cycle reduced parity check matrix}
	\label{fig:bch_63_36_ber_iregular}
\end{figure}

In Figures~\ref{fig:bch_63_45_ber_iregular} and~\ref{fig:bch_63_36_ber_iregular}, we provide the bit-error-rate for a BCH code with $N=63$ for a cycle reduced parity check matrix \cite{cycle_reduce}. For BCH(63,45) and BCH(63,36) we get an improvement up to $0.6{\rm dB}$ and $1.0{\rm dB}$, respectively. This observation shows that the method with soft Tanner graph is capable to improve the performance of standard BP even for reduced cycle parity check matrices. Thus answering in the affirmative the uncertainty in \cite{nachmani} regarding the performance of the neural decoder on a cycle reduced parity check matrix. The importance of this finding is that it enables a further improvement in the decoding performance, as BP (both standard BP and the new parameterized BP algorithm) yields lower error rate for sparser parity check matrices.

\subsubsection{BER For BCH With $N=127$}
\begin{figure}[thpb]
	\centering	\includegraphics[width=1.1\linewidth]{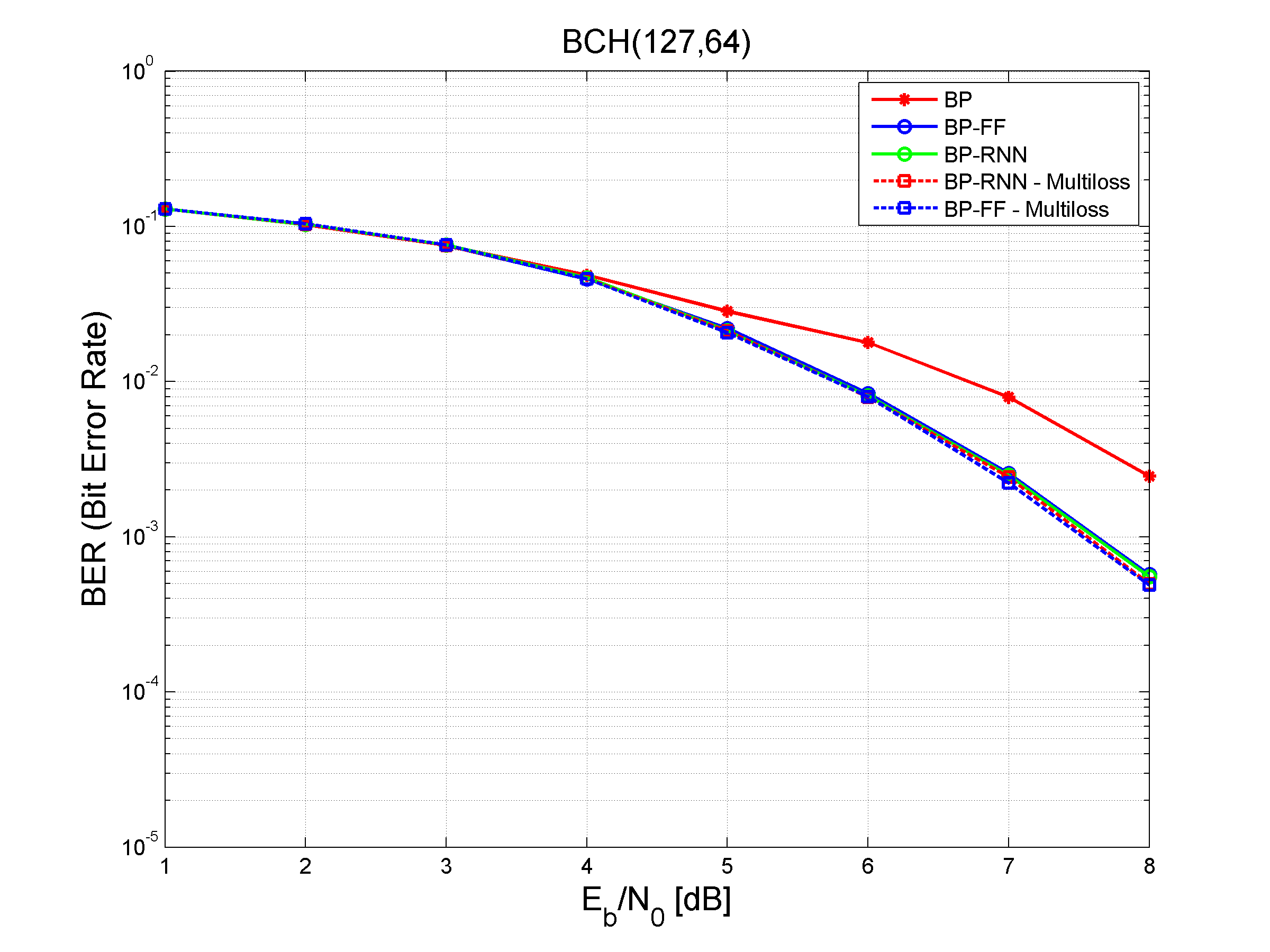}
	\caption{BER results for BCH(127,64) code trained with regular parity check matrix}
	\label{fig:bch_127_64_ber_regular}
\end{figure}

\hfill \break \newline Figure~\ref{fig:bch_127_64_ber_regular}, we provide the bit-error-rate for BCH code with $N=127$ for regular parity check matrix based on \cite{parity_g}. As can be seen from the figure, for a regular parity check matrix the BP-RNN and BP-FF decoders obtains an improvement of up to $1.0{\rm dB}$ over the BP, but BP-RNN decoder use less parameters than BP-FF.

\begin{figure}[thpb]
	\centering	\includegraphics[width=1.1\linewidth]{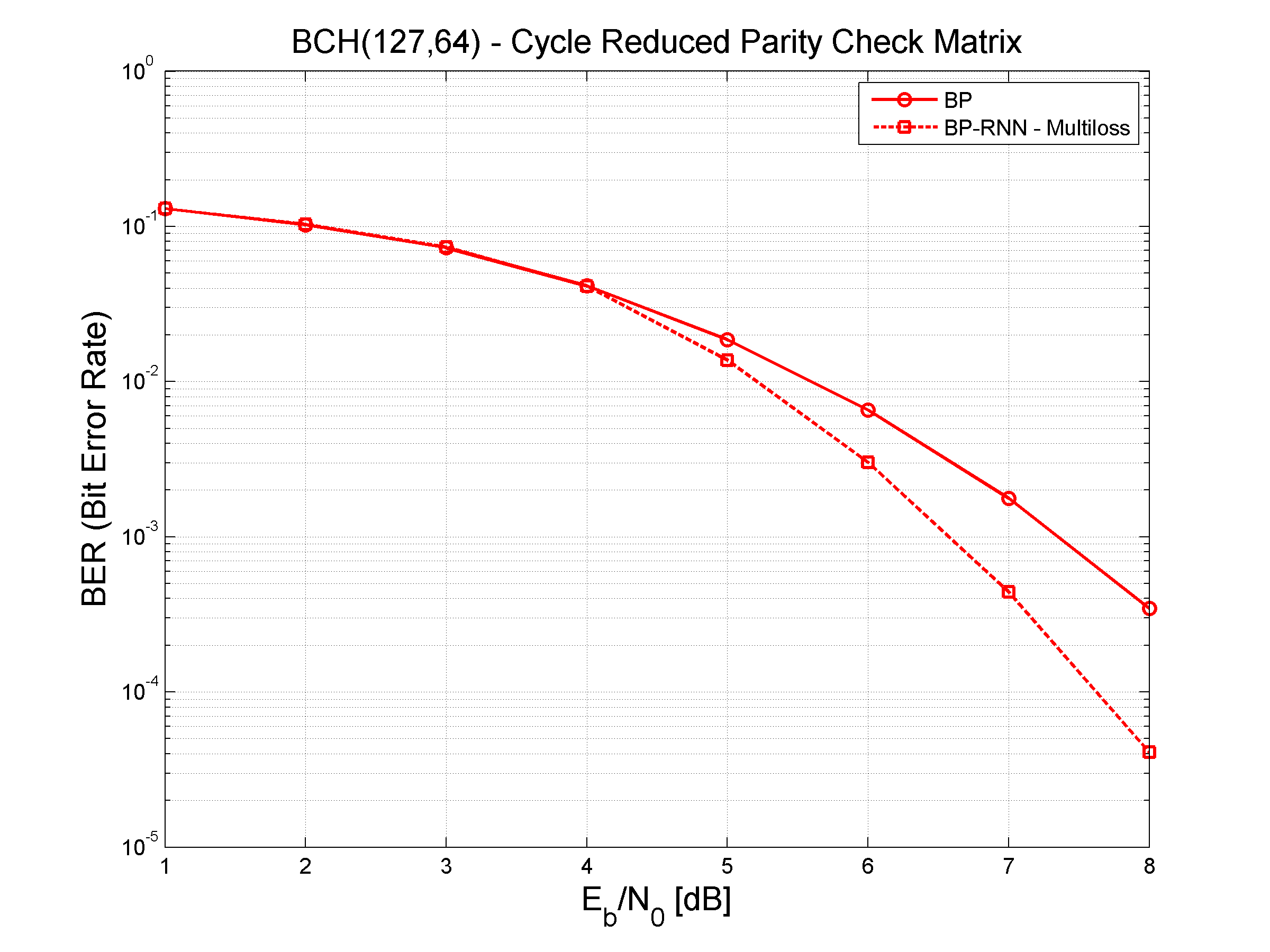}
	\caption{BER results for BCH(127,64) code trained with cycle reduced parity check matrix}
	\label{fig:bch_127_64_ber_iregular}
\end{figure}

\begin{figure}[thpb]
	\centering	\includegraphics[width=1.1\linewidth]{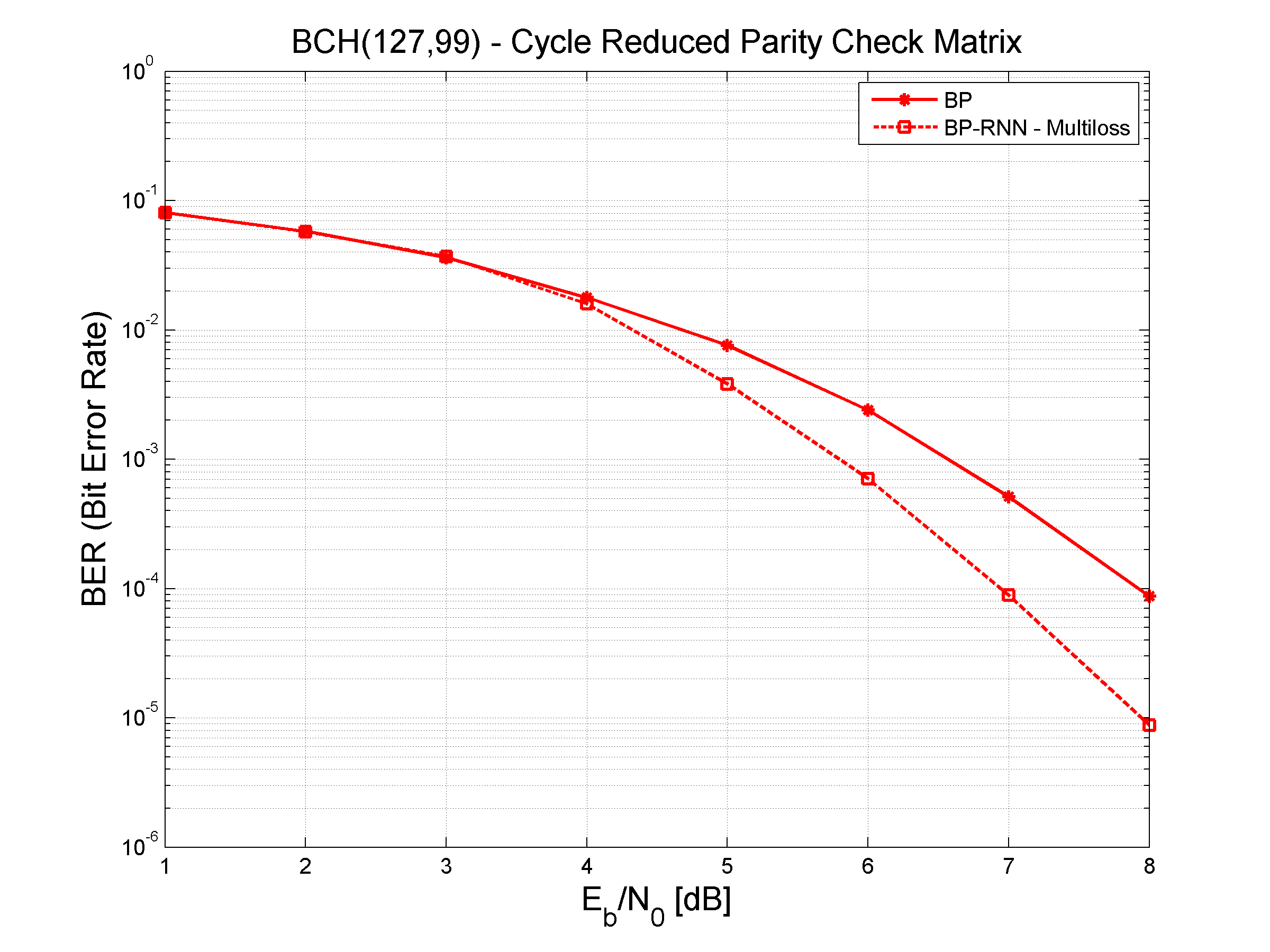}
	\caption{BER results for BCH(127,99) code trained with cycle reduced parity check matrix}
	\label{fig:bch_127_99_ber_iregular}
\end{figure}

In Figures~\ref{fig:bch_127_64_ber_iregular}, ~\ref{fig:bch_127_99_ber_iregular} we provide the bit-error-rate for BCH code with $N=127$ for cycle reduced parity check matrix based on \cite{cycle_reduce}. For BCH(127,64) and BCH(127,99) we get an improvement up to $0.9{\rm dB}$ and $1.0{\rm dB}$ respectively.

\subsection{mRRD-RNN} 
In this Section we provide the bit error rate results for a BCH(63,36) code represented by a cycle reduced parity check matrix based on \cite{cycle_reduce}. In all experiments we use the soft Tanner graph trained using the BP-RNN with multiloss architecture and an unfold of 5, which corresponds to 5 BP iterations. The parameters of the mRRD-RNN are as follows. We use 2 iterations for each ${\rm BP}_{i,j}$ block in Figure~\ref{fig:mrrd_diag}, a value of $m=1,3,5$, denoted in the following by mRRD-RNN($m$), and a value of $c=30$.

\begin{figure}[thpb]
	\centering	\includegraphics[width=1.1\linewidth]{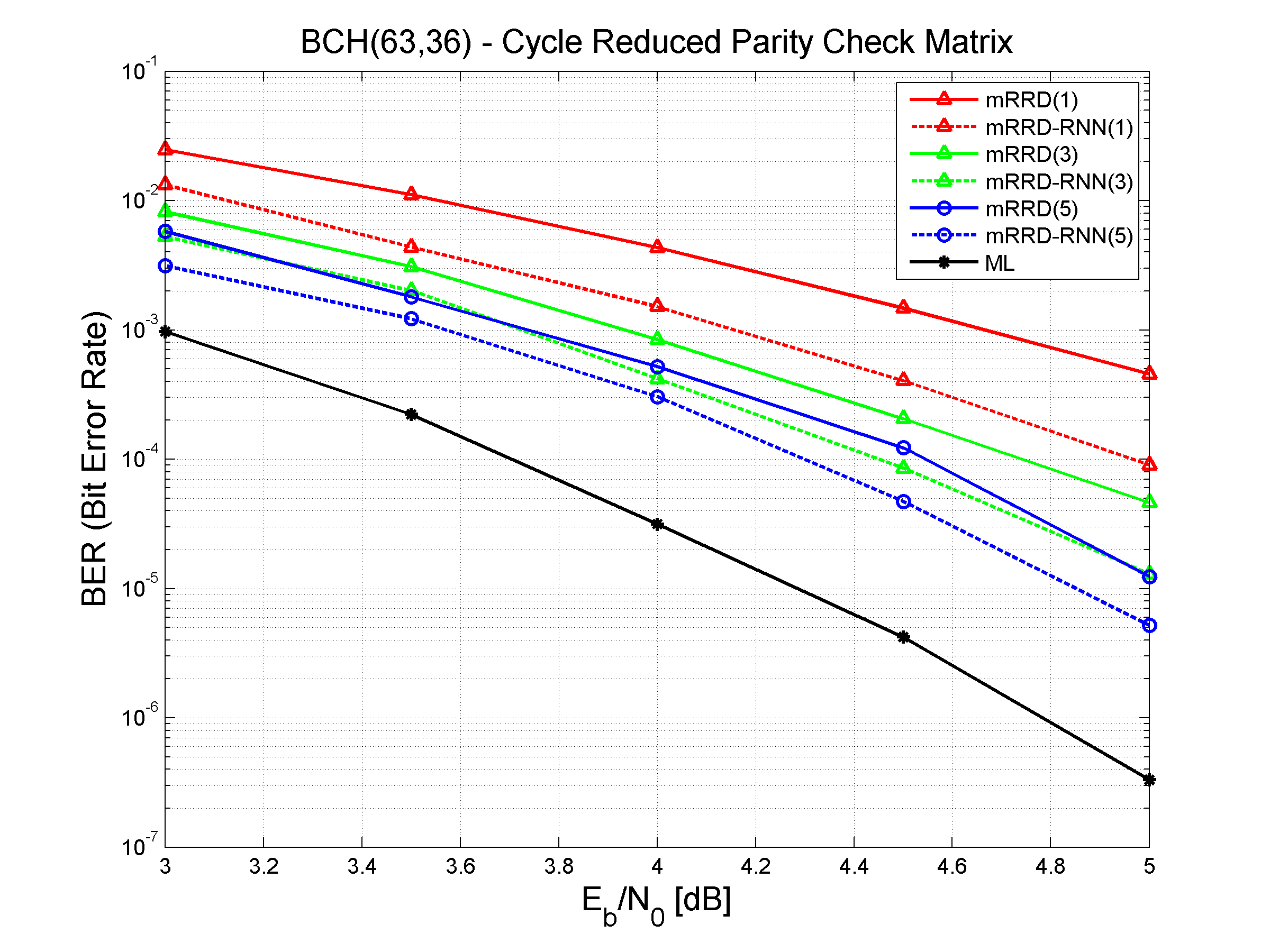}
	\caption{mRRD-RNN BER results for BCH(63,36) code}
	\label{fig:mrrd_ber}
\end{figure}

In Figure~\ref{fig:mrrd_ber} we present the bit error rate for mRRD-RNN(1), mRRD-RNN(3) and mRRD-RNN(5). As can be seen, we achieve improvements of $0.6$dB, $0.3$dB and $0.2$dB in the respective decoders. Hence, the mRRD-RNN decoder can improve on the plain mRRD decoder. Also note that there is a gap of just $0.6$dB from the optimal maximum likelihood decoder, the performance of which was estimated using the implementation of \cite{Boutros} based on the OSD algorithm \cite{fossOSD}.

\begin{figure}[thpb]
	\centering	\includegraphics[width=1.1\linewidth]{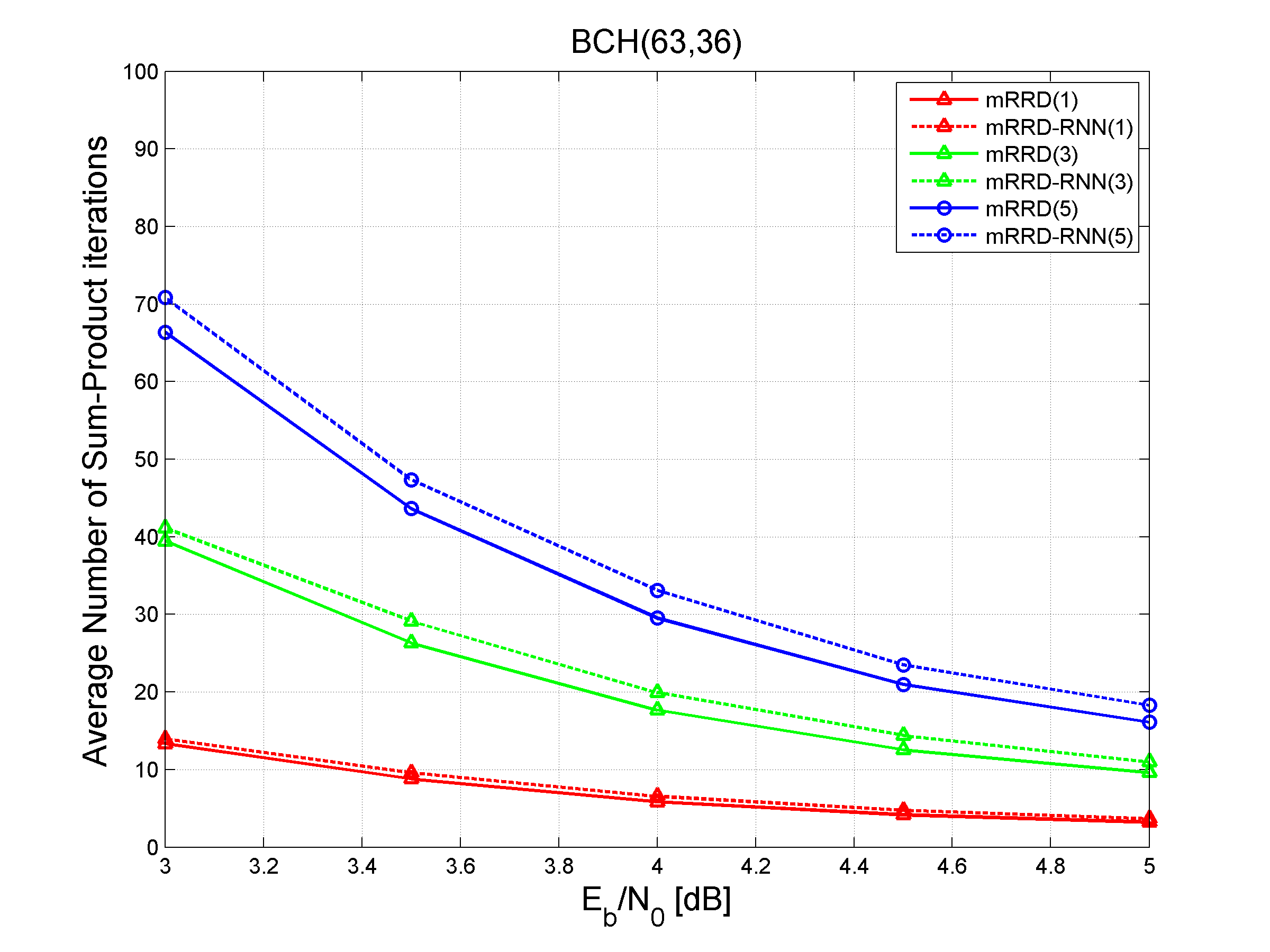}
	\caption{mRRD-RNN complexity results for BCH(63,36)}
	\label{fig:mrrd_comp}
\end{figure}

Figure~\ref{fig:mrrd_comp} compares the average number of BP iterations for the various decoders using plain mRRD and mRRD-RNN. As can be seen, there is a small increase in the complexity of up to 8\% when using the RNN decoder. However, overall, with the RNN decoder one can achieve the same error rate with a significantly smaller computational complexity due to the reduction in the required value of $m$.

\section{Conclusion} 
We introduced an RNN architecture for decoding linear block codes. This architecture yields comparable results to the feed forward architecture in \cite{nachmani} with less parameters. Furthermore, we showed that the neural network decoder improves on standard BP even for cycle reduced parity check matrices, with improvements of up to $1.0{\rm dB}$ in the SNR. We also showed performance improvement of the mRRD algorithm with the new RNN architecture. We regard this work as a further step towards the design of deep neural network-based decoding algorithms.
 
Our future work includes possible improvements in the performance by exploring new neural network architectures. Moreover, we will investigate end-to-end learning of the mRRD algorithm (i.e. learning graph with permutation), and fine tune the parameters of the mRRD-RNN algorithm. Finally, we are currently considering an extension of this work where the weights of the RNN are quantized in order to further reduce the number of free parameters. It has been shown in the past \cite{Soudry,xnor_net} that in various applications the loss in performance incurred by weight quantization can be small if this quantization is performed properly.

\addtolength{\textheight}{-12cm}   




\section*{ACKNOWLEDGMENT}

We thank Jacob Goldberger for his comments on our work, Johannes Van Wonterghem and Joseph J. Boutros for making their OSD software available to us. We also thank Ethan Shiloh, Ilia Shulman and Ilan Dimnik for helpful discussion and their support to this research, and Gianluigi Liva for sharing with us his OSD Matlab package.

This research was supported by the Israel Science Foundation, grant no. 1082/13. The Tesla K40c used for this research was donated by the NVIDIA Corporation.


\bibliographystyle{IEEEtran}
\bibliography{example_paper}

\end{document}